\documentclass[12pt,preprint]{aastex}

\newcommand{\HI}{H~{\sc i}}

\newcommand{\kms}{\hbox{km~s$^{-1}$}}
\newcommand{\cmsq}{\hbox{cm$^{-2}$}}

\newcommand{\lumin}{\hbox{erg~s$^{-1}$}}

\newcommand{\nh}{\hbox{${N}_{\rm H}$}}
\newcommand{\msun}{\hbox{${M}_{\odot}$}}

\newcommand{\rtwo}{$r_{200}$}

\newcommand{\be}{\begin{equation}}
\newcommand{\ee}{\end{equation}}
\newcommand{\ba}{\begin{eqnarray}}
\newcommand{\ea}{\end{eqnarray}}
\newcommand{\chandra}{{\emph{Chandra}}}

\newcommand{\xmm}{\emph{XMM-Newton}}

\newcommand{\simgt}{\lower 2pt \hbox{$\, \buildrel {\scriptstyle >}\over {\scriptstyle\sim}\,$}}
\newcommand{\simlt}{\lower 2pt \hbox{$\, \buildrel {\scriptstyle <}\over {\scriptstyle\sim}\,$}}
\newcommand{\ls}{\lower 2pt \hbox{$\;\scriptscriptstyle \buildrel<\over\sim\;$}}
\newcommand{\gs}{\lower 2pt \hbox{$\;\scriptscriptstyle \buildrel>\over\sim\;$}}

\newcommand{\marc}{$^{\prime}\!\!.$}

\newcommand{\ugc}{UGC~12591}

\begin{document}

\def\arcsec{$^{\prime\prime}$}
\def\arcmin{$^{\prime}$}
\def\degr{$^{\circ}$}

\title{\emph{XMM-Newton} Detects a Hot Gaseous Halo in the Fastest Rotating Spiral Galaxy UGC 12591}

\author{Xinyu Dai\altaffilmark{1}, Michael E. Anderson\altaffilmark{2}, Joel N. Bregman\altaffilmark{2}, and Jon M. Miller\altaffilmark{2}}

\altaffiltext{1}{Homer L.\ Dodge Department of Physics and Astronomy, University of Oklahoma, Norman, OK 73019, xdai@ou.edu}
\altaffiltext{2}{Department of Astronomy, University of Michigan, Ann Arbor, MI 48109}

\begin{abstract}
    We present our \emph{XMM-Newton} observation of the fastest rotating spiral galaxy UGC~12591.  We detect hot gas halo emission out to 110~kpc from the galaxy center, and constrain the halo gas mass to be smaller than $3.5\times10^{11} \msun$.  We also measure the temperature of the hot gas as $T=0.64\pm0.03$~keV.  Combining our X-ray constraints and the near-infrared and radio measurements in the literature, we find a baryon mass fraction of 0.03--0.04 in UGC~12591, suggesting a missing baryon mass of 75\% compared with the cosmological mean value.  Combined with another recent measurement in NGC~1961, the result strongly argues that the majority of missing baryons in spiral galaxies does not reside in their hot halos.
We also find that UGC~12591 lies significantly below the baryonic Tully-Fisher relationship.
Finally, we find that the baryon fractions of massive spiral galaxies are similar to those of galaxy groups with similar masses, indicating that the baryon loss is ultimately controlled by the gravitational potential well.
The cooling radius of this gas halo is small, similar to NGC~1961, which argues that the majority of stellar mass of this galaxy is not assembled as a result of cooling of this gas halo.
\end{abstract}

\keywords{galaxies: halos --- galaxies: individual (UGC 12591) --- X-rays: galaxies}

\section{Introduction}

Observations show that nearby galaxies are missing most of their baryons (e.g., Hoekstra et al.\ 2005; Heymans et al.\ 2006; Mandelbaum et al.\ 2006; Gavazzi et al.\ 2007; Jiang \& Kochanek 2007; Bregman 2007) when compared to the cosmological baryon to matter ratio (e.g., $f_b = 0.171\pm0.009$ from WMAP, Dunkley et al.\ 2009).
For example, the Milky Way is missing two thirds of its baryon allotment (e.g., Sakamoto et al.\ 2003) and less massive galaxies have retained less than 10\% of their baryons (e.g., Corbelli 2003; Walker et al.\ 2007).
This situation is confirmed in other galaxies through a variety of methods (e.g., Hoekstra et al. 2005; McGaugh 2007).
However, most of these studies do not include the baryon mass in the hot gas halo of galaxies, and it is possible that the majority of the missing baryons in galaxies actually resides in their hot has halos based on theoretical predictions (e.g., White \& Frenk 1991; Sommer-Larsen 2006; Fukugita \& Peebles 2006; also see the review by Benson 2010).  This component can be difficult to detect for spiral galaxies due to its faintness, especially if the gas density profile is flat.
As another possibility, the missing baryons from galaxies may have escaped from the potential wells of the galaxies but reside in their parent groups or clusters (e.g., Humphrey et al.\ 2011).
Finally, the missing baryons can be in the form of warm-hot intergalactic medium (Cen \& Okstiker 1999; 2006).
Deep X-ray observations are needed to distinguish these different scenarios.

The situation is different for rich galaxy clusters,
where the gas mass dominates the baryon content, and the baryon fractions in these massive systems are close to the cosmological value after combining the gas and stellar baryon contributions (e.g., Vikhlinin et al.\ 2006; Allen et al.\ 2008).
The measurements of the baryon fraction in different systems suggest that the fraction depends on the dynamical mass of the systems: rich clusters retain their cosmological allotment 
of baryons, while galaxies are baryon-poor.  
We summarize the situation in Dai et al.\ (2010) by combining the archival data points reported in the literature and our stacking analysis result using the ROSAT All-Sky Survey data of 4,000 nearby galaxy groups and clusters (Dai et al.\ 2007).
We find that the baryon fractions from dwarf galaxies to rich galaxy clusters can be fit by a broken power-law model with the break at the circular velocity of $V_c \sim 440~\kms$.  The scatter of the fractions about the mean relation is small considering the huge dynamic range of the systems.  Further examining the relation, we find that the baryon fractions are similar for different systems with similar total masses but different compositions. For example, the baryon fractions of poor galaxy groups, where the baryon mass is still dominated by the gas mass, are close to those of massive galaxies, where the baryon mass is dominated by the stellar mass.  Such a coincidence is puzzling considering the differences between their mass compositions and energy feedback mechanisms.

To test whether the missing baryons reside in galaxy halos and further constrain the baryon loss in different mass scales, we focus on the massive galaxy \ugc. 
\ugc\ is a spiral galaxy with the largest measured rotational velocity to date (466-500 km/sec, Giovanelli et al. 1986; Paturel 2003). To appreciate this
galaxy, its optical-IR luminosity is nine times that of M31. Nine big spirals were in a single group is richer
than Hickson groups and is about half of the Fornax cluster, but much more compact.
In this paper, we combine our \xmm\ observation of \ugc\ with 2MASS and other data to determine the baryon fraction and the composition of this massive spiral galaxy.
Throughout the paper, we use the cosmological parameters from WMAP with $H_0 = 72~\rm{km~s^{-1}~Mpc^{-1}}$, $\Omega_{\rm m} = 0.26$, and $\Omega_{\Lambda}= 0.74$ (Dunkley et al.\ 2009).

\section{XMM-Newton Observation and Data Reduction}

\begin{deluxetable}{cccccccccc}
\tabletypesize{\scriptsize}
\tablecolumns{10}
\tablewidth{0pt}
\tablecaption{\xmm\ Observation of \ugc\label{tab:obs}}
\tablehead{
\colhead{Sequence} & \colhead{Observation} & \colhead{CCD} & \colhead{Total} & \colhead{Effective} & \colhead{Soft Flare} & \colhead{Hard Flare} 
\\
\colhead{Number     }     & \colhead{Time} & \colhead{} & \colhead{Exposure} & \colhead{Exposure} & \colhead{Filters} & \colhead{Filters}
\\
\colhead{           }     & \colhead{    } & \colhead{} & \colhead{(ks)    } & \colhead{(ks)    } & \colhead{(cnt s$^{-1}$)} & \colhead{(cnt s$^{-1}$)}
}
\startdata
0553870101 & 2008-12-15 & PN   & 80 & 31.2 & $> 0.9$ in 0.6--1.4 keV & $> 0.4$ in 10--12 keV \\
           &            & MOS1 & 80 & 50.0 & $> 0.5$ in 0.6--1.4 keV & $> 0.2$ in 10--12 keV \\
           &            & MOS2 & 80 & 46.0 & $> 0.5$ in 0.6--1.4 keV & $> 0.2$ in 10--12 keV \\
\enddata
\end{deluxetable}

\begin{figure}
   \epsscale{0.31}
   \plotone{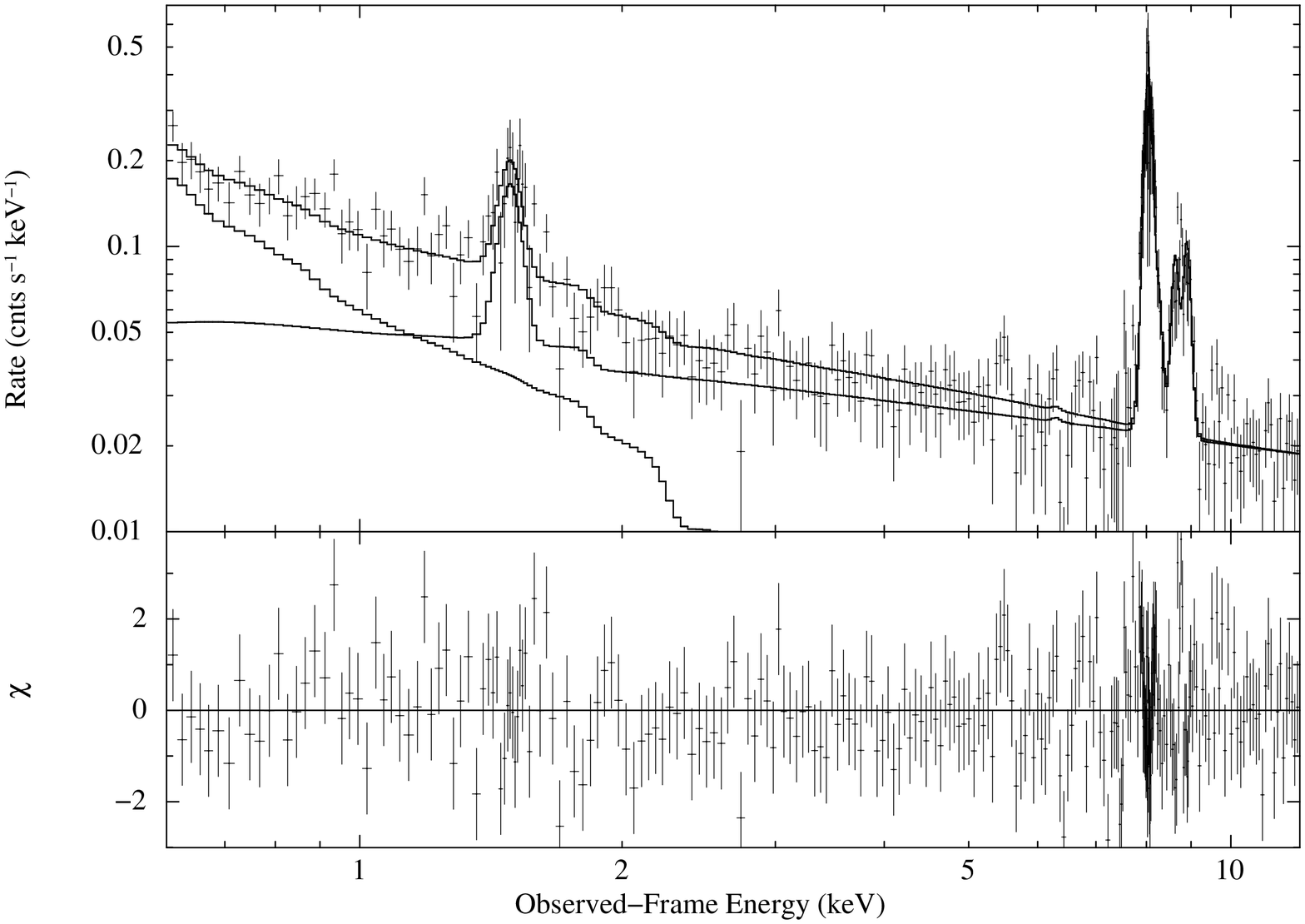}
   \plotone{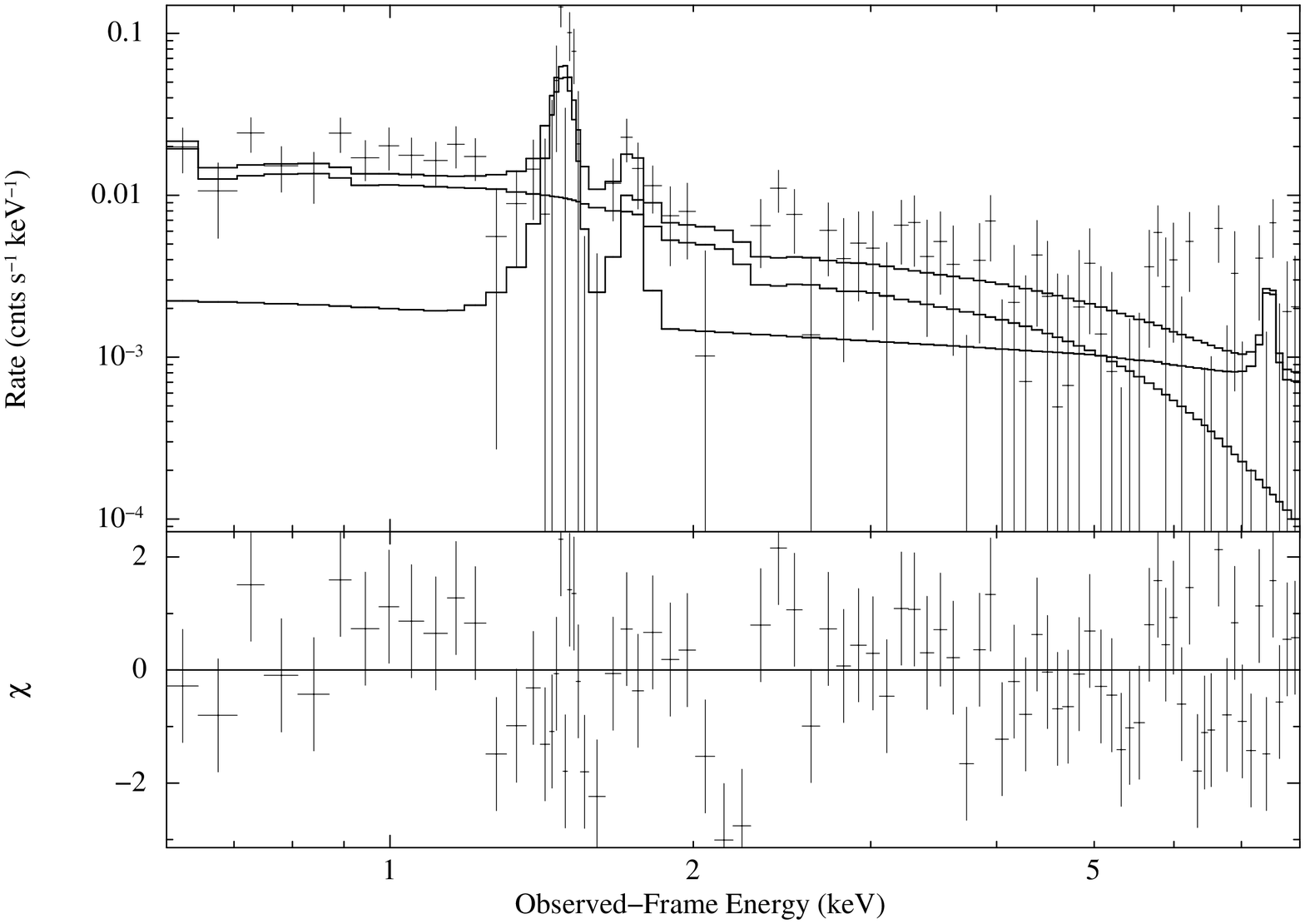}
   \plotone{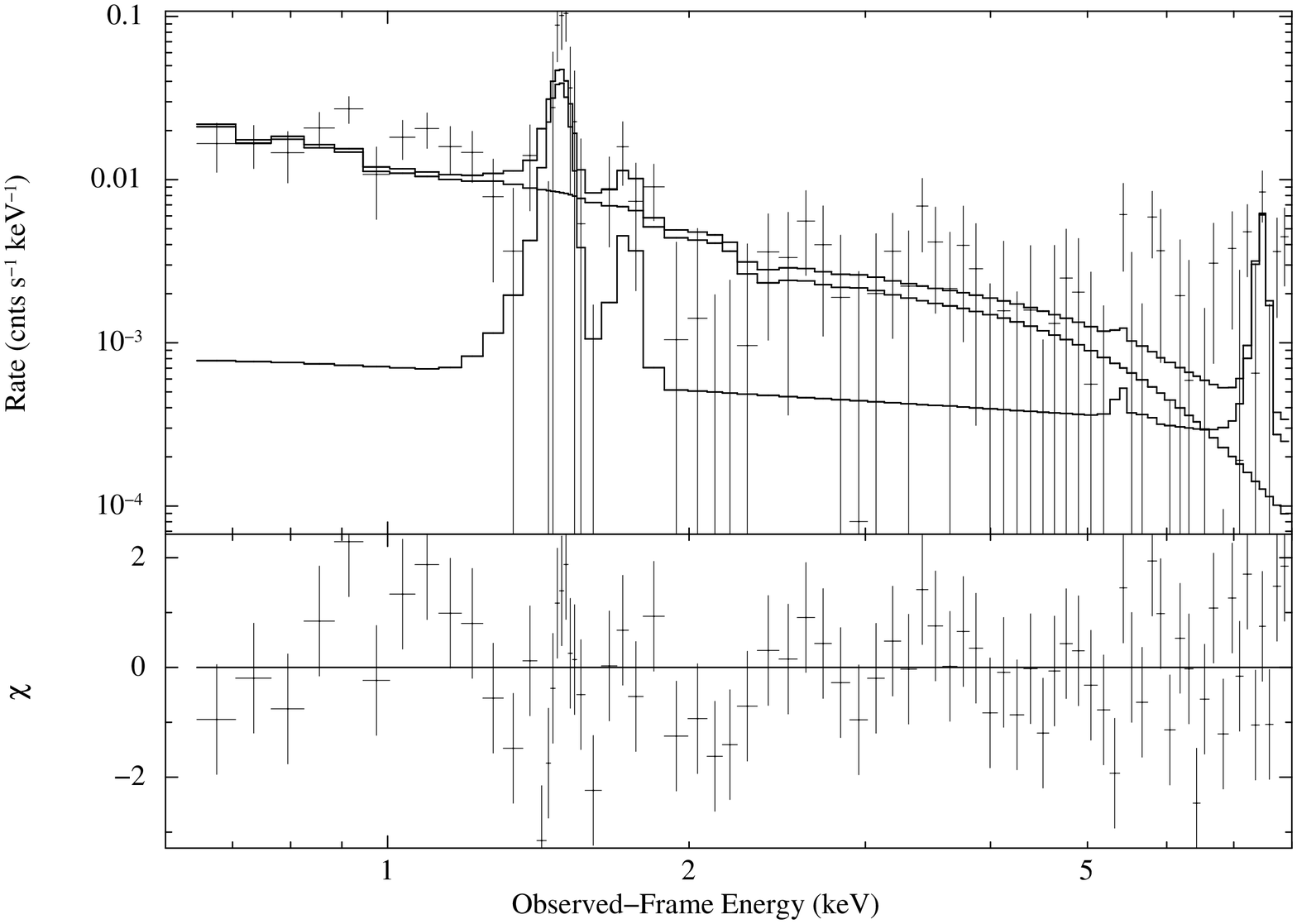}
   \caption{\xmm\ PN and MOS spectra extracted from an outer annulus background region.  We fit the spectra using the OFOV spectra as backgrounds, and modeled the spectra as a combination of photon, quiescent soft proton, and residual metal emission line induced backgrounds. \label{fig:bkgspec}}
\end{figure}

We observed \ugc\ with the \xmm\ X-ray Observatory (Jansen et al.\ 2001) on December 15, 2008 for 80 ks. 
The observation log is listed in Table~\ref{tab:obs}.
We reprocessed the PN and MOS data using the \verb+SAS9.0.0+ software tools \verb+epchain+ and \verb+emchain+, and filtered the events with the patterns $\le 4$ and $\le 12$ for the PN and MOS chips, respectively.
We filtered background flares by excluding the intervals with background count rates of $CR > 0.4$ cnt~s$^{-1}$ and  $CR > 0.2$ cnt~s$^{-1}$ in the 10--12 keV band in PN and MOS observations, respectively, following the standard suggestion from SAS.
We also applied a low energy flare filter to exclude flares in the 0.6--1.4 keV band (Table~\ref{tab:obs}).
We obtained net exposure times of 31.2, 50.0 and 46.0 ks for PN, MOS1 and MOS2 CCDs, respectively.
We detected serendipitous sources in the field using the SAS tools, and masked the serendipitous source regions in the subsequent analysis.
The central region of \ugc\ is clearly detected in X-rays in all PN and MOS CCDs.  However, a detailed analysis of the background is needed to determine the extent to which the outer region is detected.

\subsection{Background Determination}
We estimated the PN and MOS backgrounds by directly fitting the background spectra (e.g., Kuntz \& Snowden 2008; Leccardi \& Molendi 2008).
Here, we briefly describe the method.  The \xmm\ background is composed of several components induced by soft protons, cosmic rays, galactic, and extra-galactic background photons (e.g., Carter \& Read 2007).
After filtering the background flares in the hard and the soft energy bands, we had removed most of the variable soft proton background.
We then extracted the spectrum from an outer annulus background region, and the remaining background components in this spectrum is from quiescent soft protons, cosmic rays, galactic, and extra-galactic background photons.
The cosmic ray background can be subtracted by comparing the surface brightnesses in the outer ring and out-of-FOV (OFOV) regions.
We can then decompose the contributions from the photon background and quiescent soft proton background through spectral analysis.
We fixed the spectral index of the extra-galactic background as $\Gamma=1.4$ and Galactic background with $T=0.197$~keV, and allowed the normalizations of these two components to vary during the spectral fits.
Figure~\ref{fig:bkgspec} shows the PN and MOS spectra of the outer ring background region, where we used the OFOV spectra as backgrounds.
The spectra are well fit by the combination of photon background, quiescent soft proton background, and residual metal emission lines.
We estimated that the ratios between the quiescent soft proton background and photon background are 0.676, 0.200, and 0.087 for PN, MOS1, and MOS2, respectively, in the 0.6--1.4 keV band.
After this step, we can scale the two components to inner regions using the corresponding vignetting profiles.

\begin{figure}
   \epsscale{1}
   \plotone{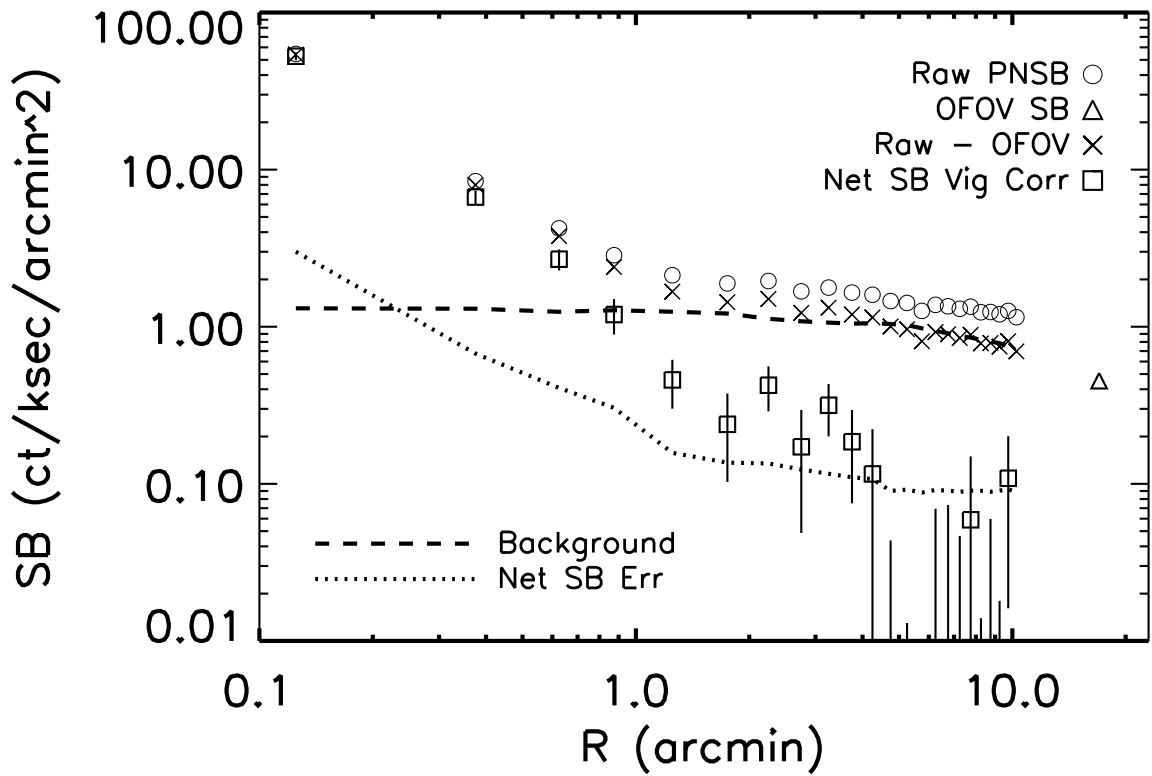}
   \caption{Surface brightness profile of \ugc\ from the \xmm\ PN observation.  The circles represent the raw surface brightness profile, and the triangle represents the OFOV surface brightness.  The cross symbols are results of subtracting the raw surface brightness profile by the OFOV surface brightness.
The dashed line is our model for the background including the contributions of photon, quiescent soft proton, and residual metal emission lines, which fit the outer regions well.  The squares and their associated error-bars denote the net surface brightness profile and its uncertainty of \ugc\ (crosses minus the dashed line) corrected for vignetting.  We also plot the uncertainties as dotted lines. \label{fig:pn}}
\end{figure}
\begin{figure}
   \epsscale{1}
   \plotone{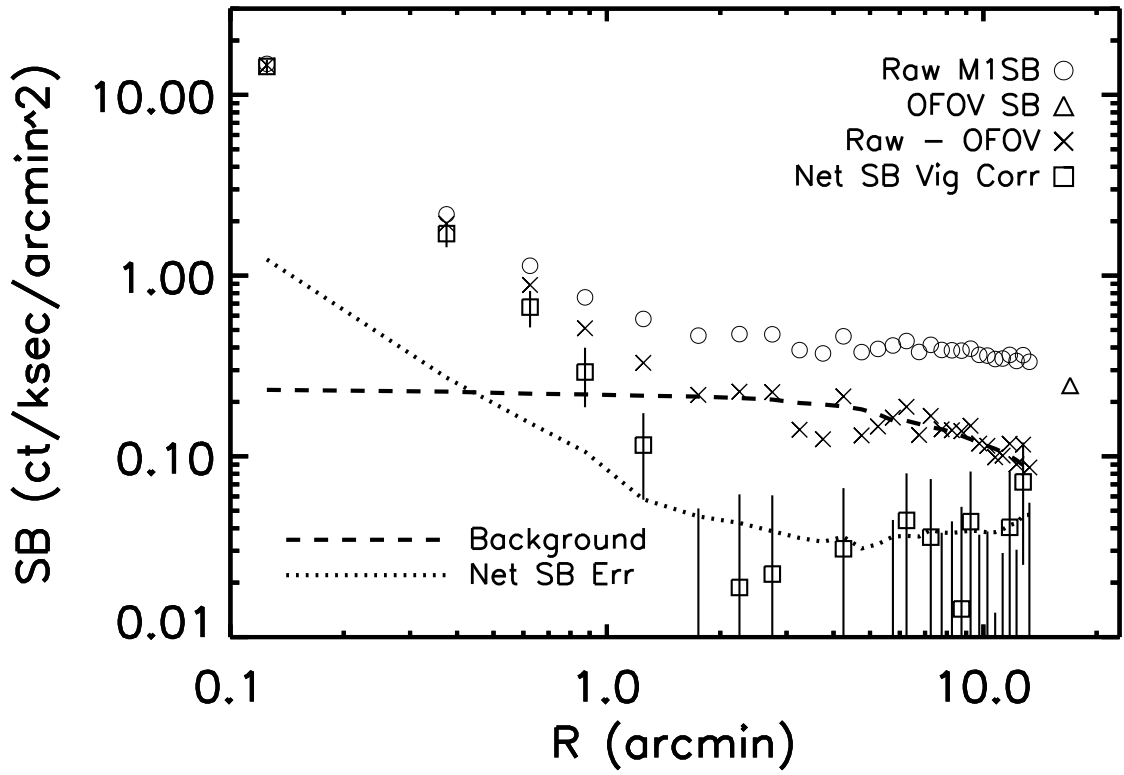}
   \caption{Surface brightness profile of \ugc\ from the \xmm\ MOS1 observation. \label{fig:mone}}
\end{figure}
\begin{figure}
   \epsscale{1}
   \plotone{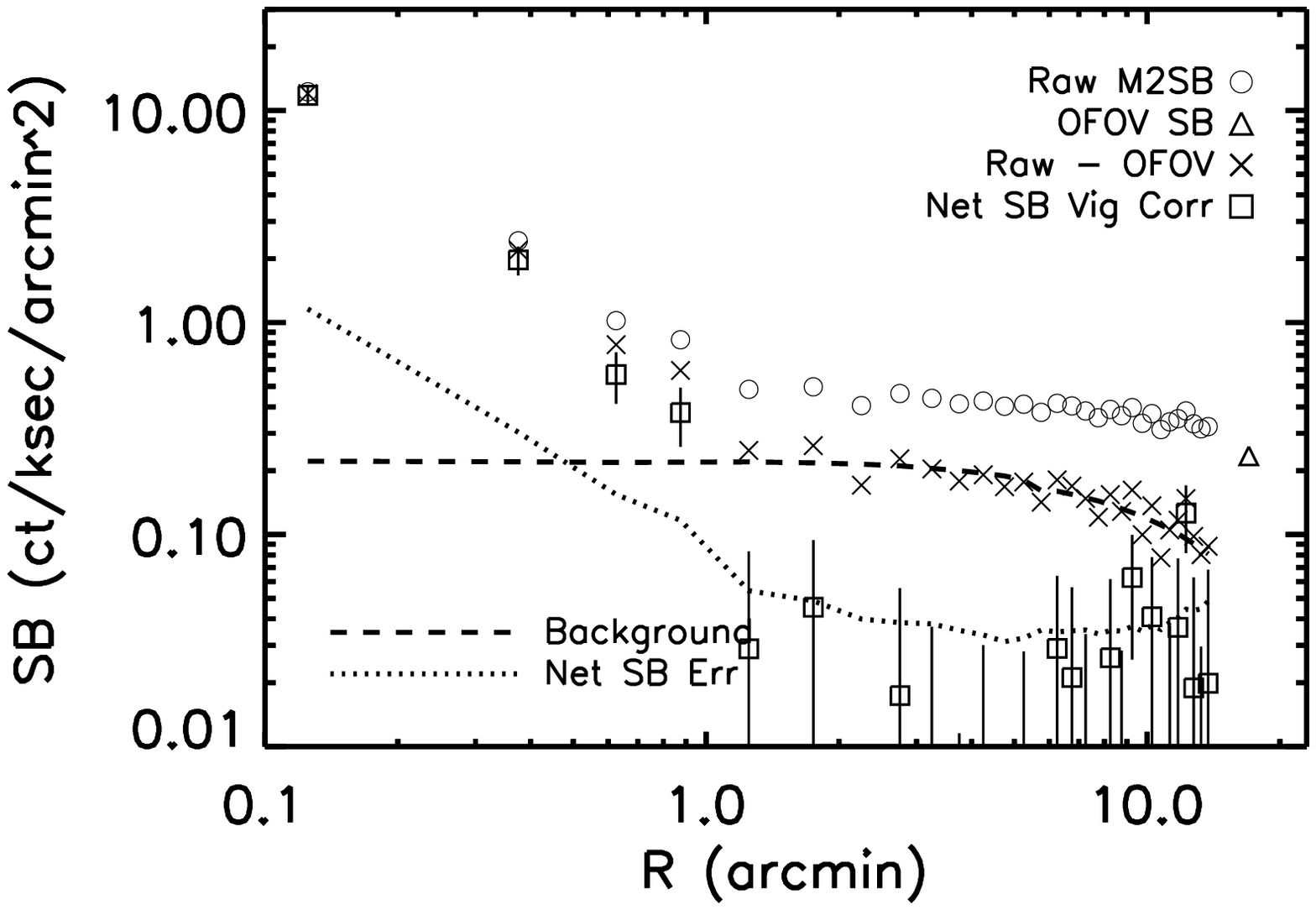}
   \caption{Surface brightness profile of \ugc\ from the \xmm\ MOS2 observation. \label{fig:mtwo}}
\end{figure}

\subsection{Surface Brightness Profile of \ugc}
We measured the surface brightness profile of \ugc\ by extracting events in consecutive annuli of 0\marc5 widths (13.6~kpc).  For the region within 1\arcmin\ from the center, we further divided the annuli using 0\marc25 widths.  We measured average exposure times within each annulus.
To increase the signal-to-noise ratio of the surface brightness profile, we measured it in the soft X-ray (0.6--1.4 keV) band.
We first subtracted the surface brightness of the OFOV region, which is un-vignetted, from the surface brightness profile.
We then subtracted the photon and non-photon induced background based on our background decompositions obtained by fitting the outer background spectra.
We plot the net surface brightness profiles and their uncertainties of \ugc\ after correcting for vignetting measured from PN and MOS CCDs, in Figures~\ref{fig:pn}--\ref{fig:mtwo}, respectively.
\ugc\ is detected out to $\sim$3\arcmin\ in the PN data and $\sim$1\marc5 in the MOS data. 
We also extracted the surface brightness profile of \ugc\ in the hard X-ray band between 2--8 keV to measure the contribution from point sources.  The hard component from point sources is detected to 1\arcmin\ and 0\marc5, respectively, in the PN and MOS data.
After scaling the 2--8 keV count rate to the 0.6--1.4 keV count rate for the point source contribution, assuming $\Gamma=1.56$ (Irwin et al.\ 2003), we found the contribution from point sources in the 0.6--1.4 keV band flux is 45\% within the central 1\arcmin\ region.

\subsection{Spectral Analysis}
\begin{figure}
   \epsscale{1}
   \plotone{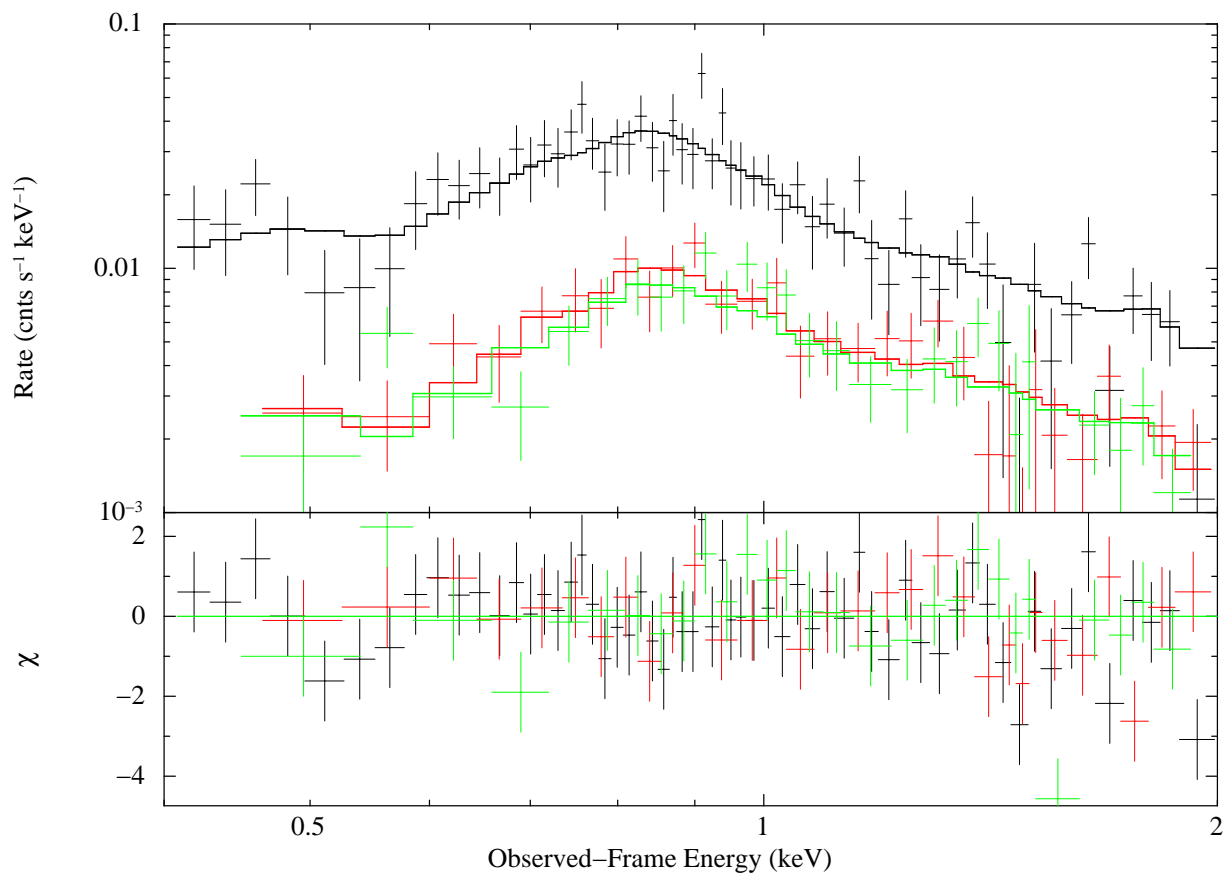}
   \caption{\xmm\ PN and MOS spectra of the central region of \ugc. We fit the spectra using a combination of a power law and an APEC model both modified by Galactic absorption and absorption at the source. \label{fig:spec}}
\end{figure}

We extracted the central spectra of \ugc\ within 50\arcsec\ from the PN and MOS data.
The background regions were chosen in a region with large off-axis angles to avoid possible contamination from the extended halo emission detected in  \ugc.
We built the \verb+rmf+ and \verb+arf+ files separately for the source and background spectra.
We modeled the source spectra by combining an APEC model for the extended gas emission and a power-law model for point sources, modified by Galactic absorption and absorption at the source.  We fixed the photon index of the power-law component as $\Gamma=1.56$ and scaled its normalization based on the 2--8 keV count rate of the central 50\arcsec\ region, since the expected gas emission (the APEC model) in this band is negligible.
We also fixed the Galactic absorption based on the value obtained by \citet{d90}.
Since our spectra cannot constrain the gas metallicity reliably, we fixed it at 0.5 solar metallicity.
The free parameters left are the temperature and normalization of the gas emission and the \nh\ column density of the absorption at the source. 
Figure~\ref{fig:spec} shows our simultaneous fit to the PN and MOS spectra using this model.  We obtained accepted fit to the spectra with $\Delta \chi^2/dof = 131.5/115$, and constrained the gas temperature as $T=0.64\pm0.03$~keV and the absorption at the source as $\nh = (5\pm2)\times10^{20}\cmsq$.  
The 0.2--10~keV luminosity for the point source component is $1.1\times10^{41} \lumin$.
At this luminosity, the central source is possibly dominated by an AGN.

\section{Mass of the Hot Gaseous Halo}
We fit the surface brightness profile in the 0.6--1.4 keV band using three components, the hot gas, AGN/binaries, and the contribution from the stellar population (including cataclysmic variables). 
The AGN/binary contribution can be measured using the surface brightness profiles from the 2--8 keV band, as described above.  We estimated the X-ray emission from the stellar population (excluding the binaries) using the stellar mass to X-ray luminosity conversion factor of Revnivtsev et al. (2008). 
This relation is calibrated for old stellar populations, and since UGC 12591 is fairly bulge-dominated, the relation should still work approximately. 
To compute the radial surface brightness profile from the stellar population in the X-ray band, we derived a K-band radial surface brightness profile from the K-band magnitudes within circles of different angular sizes for this galaxy listed in the 2MASS Extended Source Catalog. 
We used a distance modulus of 35.0 and a K-band mass-to-light ratio of 0.6 (Bell and de Jong 2001) for this galaxy, and then applied the Revnivtsev et al. (2008) conversion to derive an X-ray surface brightness profile. 
The remaining emission attributes to the hot gas in the halo. We fit the surface brightness profile of this emission using a standard $\beta$ model. 

\begin{figure}
   \epsscale{1}
   \plotone{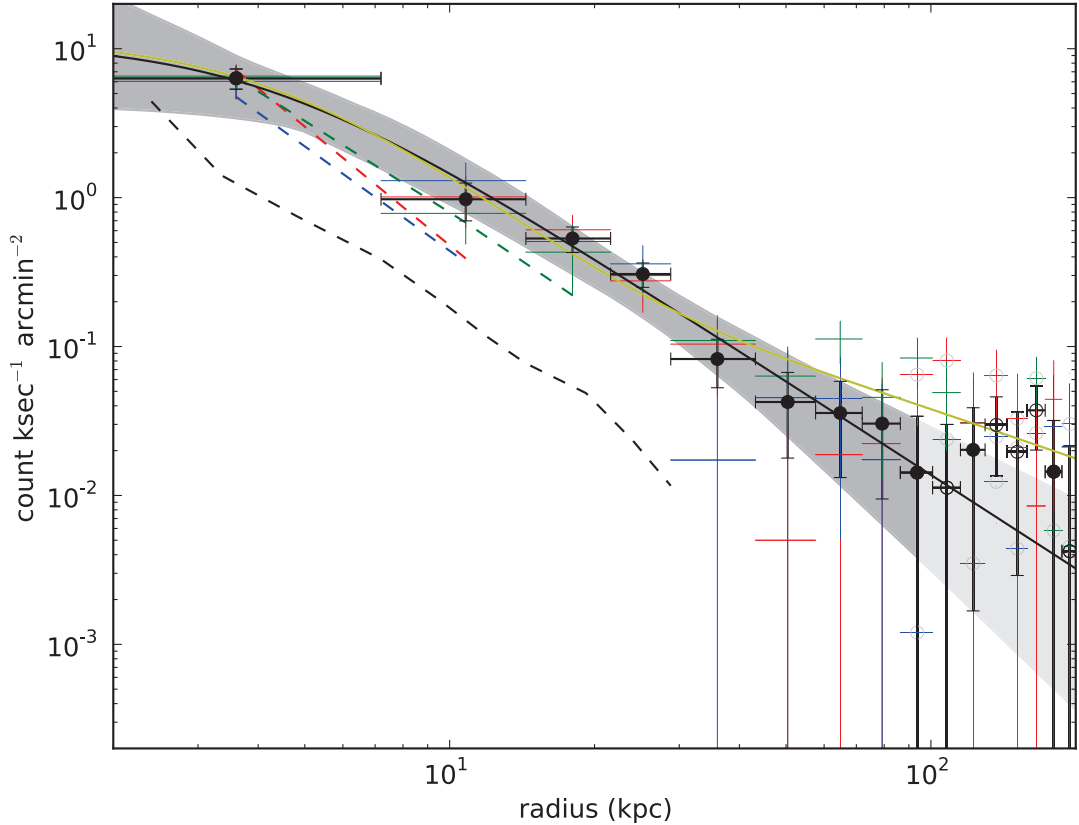}
   \caption{Background-subtracted and un-vignetted \xmm\ PN and MOS radial surface brightness profiles of the 0.6-1.4 keV emission around UGC 12591 (the open circles denote negative values). The PN emission has been multiplied by a factor of 0.26 to convert it into effective counts for the MOS detectors. 
We fit the surface brightness profile with the sum of three components, the hot gas halo, AGN/binaries, and the stellar contribution. The black dashed line is the emission from coronally active stars and cataclysmic variables in the galaxy, estimated by scaling the 2MASS K-band luminosity. The colored dashed lines are the emission from X-ray binaries/AGN, estimated by scaling from the 2-8 keV surface brightness profile. Finally, the shaded region denotes the emission from hot gas, which makes up the rest of the observed surface brightness profile. The black solid line is the best-fit $\beta$-model for the hot halo component with a $\chi^2$ of 61.1 for 45 degrees of freedom, and the green solid line shows our best fit by adding an additional flattered $\beta$ profile. \label{fig:loglog}}
\end{figure}

Figure~\ref{fig:loglog} shows the \xmm\ PN and MOS data, after vignetting correction and background subtraction, as well as the estimates of the various components we fit to the surface brightness profile. 
It is clear that out to 30 or 40 kpc the X-ray binary and stellar components are insufficient to account for all the emission, and therefore that a hot halo component is necessary. We only allowed $\beta$-models for the hot gas if they cannot be excluded at greater than 95\% confidence. This results in a narrow range of acceptable fits for the hot halo profile. The true uncertainties in the surface brightness profile are probably somewhat larger than the statistical errors, however, due to inevitable systematic errors in the flat fielding and background subtraction. 
This means that the range of formally acceptable fits for the hot halo might be a little wider than we have indicated Figure~\ref{fig:loglog}. 
Deviations from a simple $\beta$-model for the hot gas distribution are also possible, and we tested by adding a second flatter $\beta$ model component in our fits.

The acceptable fit with the highest enclosed mass has $\beta = 0.43$, $r_0 =  1.0$ kpc, and $S_0 = 45.3$ counts ksec$^{-1}$ arcmin$^{-2}$. This corresponds to a count rate within a projected radius of 50 kpc of 0.0032 counts s$^{-1}$. Assuming an APEC model with $kT = 0.64$, $Z = 0.5 Z_{\odot}$, and $N_H$ = $5\times10^{20}$ cm$^{-2}$, we constrained the mass within 50 kpc as $4.4\times10^9 M_{\odot}$ and the unabsorbed luminosity (0.6--1.4 keV) of $2.3\times10^{40}$ erg s$^{-1}$ given a distance of 100~Mpc. 
The acceptable fit with the lowest enclosed mass has $\beta = 0.69$, $r_0 = 8.6$ kpc, and $S_0 = 6.7$ counts ksec$^{-1}$ arcmin$^{-2}$. This corresponds to a count rate within a projected radius of 50 kpc of 0.0028 counts s$^{-1}$, yielding a mass of $3.9\times10^9 M_{\odot}$ and an unabsorbed luminosity of $2.3\times10^{40}$ erg s$^{-1}$. 
The best-fit profile has $\beta = 0.52$, $r_0 = 4.05$ kpc, and $S_0$ = 11.2 counts ksec$^{-1}$ arcmin$^{-2}$. The $\chi^2$ is 4.6 for 6 degrees of freedom. 
If we integrate these profiles out to 500 kpc, the fit with the highest enclosed mass contains $2.2\times10^{11} M_{\odot}$ with an unabsorbed luminosity of $3.8\times10^{40}$ erg s$^{-1}$. The fit with the lowest enclosed mass contains $4.4\times 10^{10} M_{\odot}$ with an unabsorbed luminosity of $3.2\times10^{40}$ erg s$^{-1}$. 

We also examined the possibility of a higher-entropy halo as predicted by many simulations (Maller and Bullock 2004; Kaufmann et al. 2009; Crain et al. 2010; Guedes et al. 2011). Such a halo would have a flatter density profile than the $\beta \sim 0.5$ models we find above, and it would therefore be both more difficult to detect in emission and also more massive than a lower-entropy halo. 
As in Anderson and Bregman (2011), we chose to model a flattened profile with a two-component fit to the data. The ``flattened'' component is a $\beta$-model with fixed $\beta = 0.35$ and $r_0 = 50$ kpc, but with free normalization, and the other component is a more concentrated $\beta$-model with all three parameters free, used to model the emission at smaller radii. For this galaxy, there is very little statistical space left to add a flattened profile, so the most mass that can be included by adding a flattened component is $3.5\times10^{11} M_{\odot}$. 
As before, however, we caution that this statistical constraint depends on understanding all the systematic uncertainties perfectly, especially the background in \xmm\ PN/MOS CCDs. 
These profiles can be independently tested in absorption profiles instead of emission due with the linear dependence on density in absorption.  
In addition, we note that the metallicity assumed can be another uncertainty in our analysis since the total mass will depend on the metallicity.

\section{Discussion}
\begin{deluxetable}{lccccccccc}
\tabletypesize{\scriptsize}
\tablecolumns{10}
\tablewidth{0pt}
\tablecaption{Gravitational and Baryon Mass Components in \ugc \label{tab:mass}}
\tablehead{
\colhead{Component} &
\colhead{within 50~kpc radius} &
\colhead{within 500~kpc radius} 
}
\startdata
Stellar           & $45\pm10$     & $45\pm10$ \\
Cold Gas          & 0.7           & 0.7       \\
Hot Gas           & $0.41\pm0.03$ & 13        \\
Including Flattened Hot Gas & 0.55       & $\ls 35$  \\
Gravitational     & 270           & 1900      \\
Stellar-to-gas ratio $r_{sg}$ & 39         & 3.3       \\
$r_{sg}$ including the flattened gas & 33 & 3.3--1.3 \\
Baryon Fraction $f_b$  & 0.17          & 0.03      \\
$f_b$ including the flattened gas & 0.17 & 0.03--0.04 \\
\enddata
\tablecomments{The masses are in the unit of $10^{10} \msun$.  The \rtwo\ is at $\simeq 550$~kpc.}
\end{deluxetable}

\subsection{Baryon Mass Components in \ugc}
We list the various baryon mass components in \ugc\ in Table~\ref{tab:mass}.
We have constrained the hot gas mass of \ugc\ using the \xmm\ observation as $(4.1\pm0.3)\times10^{9} \msun$ within 50~kpc with an average temperature of $T=0.64\pm0.03$~keV. We have also constrained the hot gas mass of $1.3\times10^{11} \msun$ within 500~kpc regions using our best fit $\beta$ model, and the hot gas mass is below $3.5\times10^{11} \msun$ within 500~kpc even if we add another flatter $\beta$ model component in our fits.
Beside the hot gas mass, there are other baryon mass components in the galaxy including the stellar mass and cold gas mass components.
For the cold gas mass component, \citet{gi86} measure the \HI\ mass of $5.3\times10^{9} h_{72}^{-2} \msun$ from the radio data.
Assuming the \HI\ mass is 75\% of the total cold gas mass, we find that the total cold gas mass is $M_{cg} = 7.1\times10^9 \msun$, where we ignore the contribution from the molecular gas component.
We estimate the stellar mass of \ugc\ as $(4.5\pm1.0)\times10^{11} \msun$ within 29~kpc radius using its $K$ band total magnitude ($K = 8.89$~mag) from the 2MASS Extended Source Catalog and a range of mass-to-light ratio from 0.6 to 0.95 (e.g., Bell et al.\ 2003).
The 2MASS team calculates the total magnitude by integrating the surface brightness profile out to $\sim 4$ disk scale lengths from the isophotal aperture well below the $1\sigma$ noise level.
For the mass-to-light ratio, Bell et al.\ (2003) measure a value of 0.95 as the cosmic mean value.  However, since \ugc\ is a late-type galaxy and could have a lower mass-to-light ratio, we choose to use $0.78\pm0.18$ in our calculation.
We find the largest uncertainties in the baryon mass are from the systematical uncertainty in the stellar mass-to-light ratio and the flattened gas halo. Combining the two effects, we find an uncertainty of $2.4\times10^{11} \msun$ for the total baryon mass within 500~kpc.

In the central region within $\sim50$~kpc, the baryon mass is clearly dominated by the stellar mass, and the stellar-to-gas mass ratio is $r_{sg} \simeq 39$.  Using the rotational velocity of 466--500~\kms, we measure a total mass of $m_{tot} = 2.7\times10^{12} \msun$ within 50~kpc and a baryon mass fraction of $f_b \simeq 0.17$, consistent with the cosmological baryon fraction.
Out to the 500~kpc region (\rtwo $\simeq 550$~kpc), we use the gravitational mass $m_{tot} = 1.9\times10^{13} \msun$ estimated from the X-ray data, because the rotational curve is only constrained within 28$h_{72}^{-1}$~kpc \citep{gi86}, smaller than the total mass $m_{tot} = 2.7\times10^{13} \msun$ estimated using the rotational curve.  The baryon mass within 500~kpc is $m_b = 5.9\times10^{11} \msun$, and we measure a baryon fraction of $f_b \simeq 0.03$.  Considering a second flattened gas component, the baryon fraction within 500~kpc can reach to $f_b \ls 0.04$.
Since we use the smaller total mass estimate in the calculation, the baryon fraction quoted should be treated as a conservative upper limit. The stellar mass component is still more important with $r_{sg} = 3.3$ within 500~kpc, or $r_{sg} \gs 1.3$ with the additional flatter gas component.

 To summarize, combining our \xmm\ observation and the 2MASS and radio data in the literature, we have constrained that \ugc\ has lost at least 75\% of the baryons compared to the cosmological value.  The missing baryons do not reside in the hot halos for spiral galaxies.  Our result confirms the recent measurements in another giant spiral NGC~1961 using \chandra\ by Anderson \& Bregman (2011), who find that NGC~1961 has also lost 75\% of its baryon content.

\subsection{Baryonic Tully-Fisher Relationship}
\begin{figure}
   \epsscale{1}
   \plotone{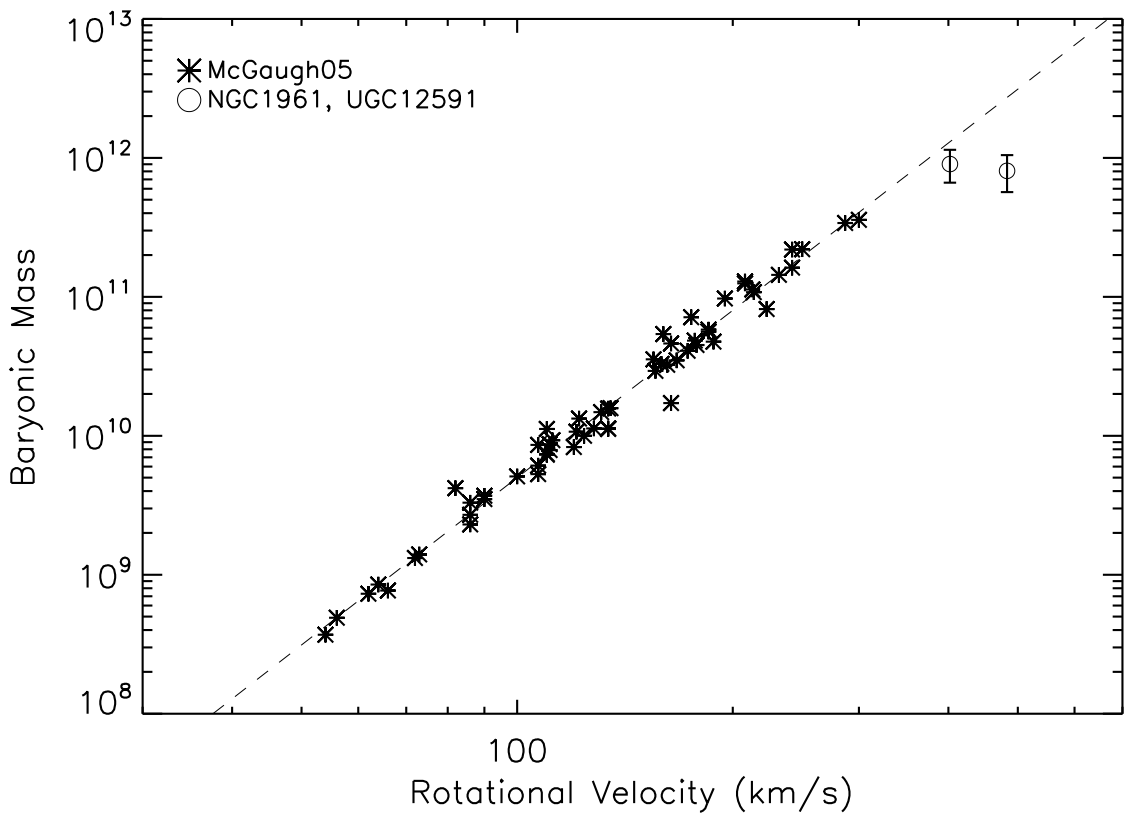}
   \caption{The baryonic Tully-Fisher relationship of McGaugh (2005) adding NGC~1961 (Anderson \& Bregman 2011) and \ugc. \label{fig:btf}}
\end{figure}
Using the fastest rotating galaxy \ugc, we are able to extend the baryonic Tully-Fisher relationship (BTF), a correlation between the baryon mass and rotational velocity of galaxies (McGaugh 2005; 2011) to the high rotational velocity regime of 500~\kms.
We plot \ugc\ in the BTF diagram together with the galaxies in McGaugh (2005) and the other massive galaxy NGC1961 (Anderson \& Bregman 2011) with a rotational velocity of 402~\kms\ in Figure~\ref{fig:btf}.
Anderson \& Bregman (2011) find that NGC1961 deviates slightly from the linear fits to BTF.
However, the authors caution that the offset can be caused by systematic uncertainties.
Indeed, assuming a $K$ band mass-to-light of 0.95, NGC~1961 would be on the BTF relation.
Here, \ugc\ provides another challenge to the linear BTF relation, which predicts a baryon mass of $2.2\times10^{12} \msun$ for \ugc, whereas we measure a baryon mass in the range of 5.9--8.1$\times10^{11} \msun$ with an uncertainty of $2.4\times10^{11} \msun$.  Thus, \ugc\ is $6\sigma$ below the BTF relation.  If the offset from the BTF relation is caused by the uncertainties in the $K$ band mass-to-light ratio, a ratio of 3.1 is needed to put \ugc\ on the BTF relation, which is extremely unlikely (e.g., Bell et al.\ 2003). 
Thus, it is possible that the BTF relation turns over for massive galaxies with $v_c \gs 400 \kms$.
However, measurements from a larger sample of massive spiral galaxies are needed to confirm this result.

\subsection{Overall Relationship of Baryon Fractions with Total Mass}
\begin{figure}
   \epsscale{1}
   \plotone{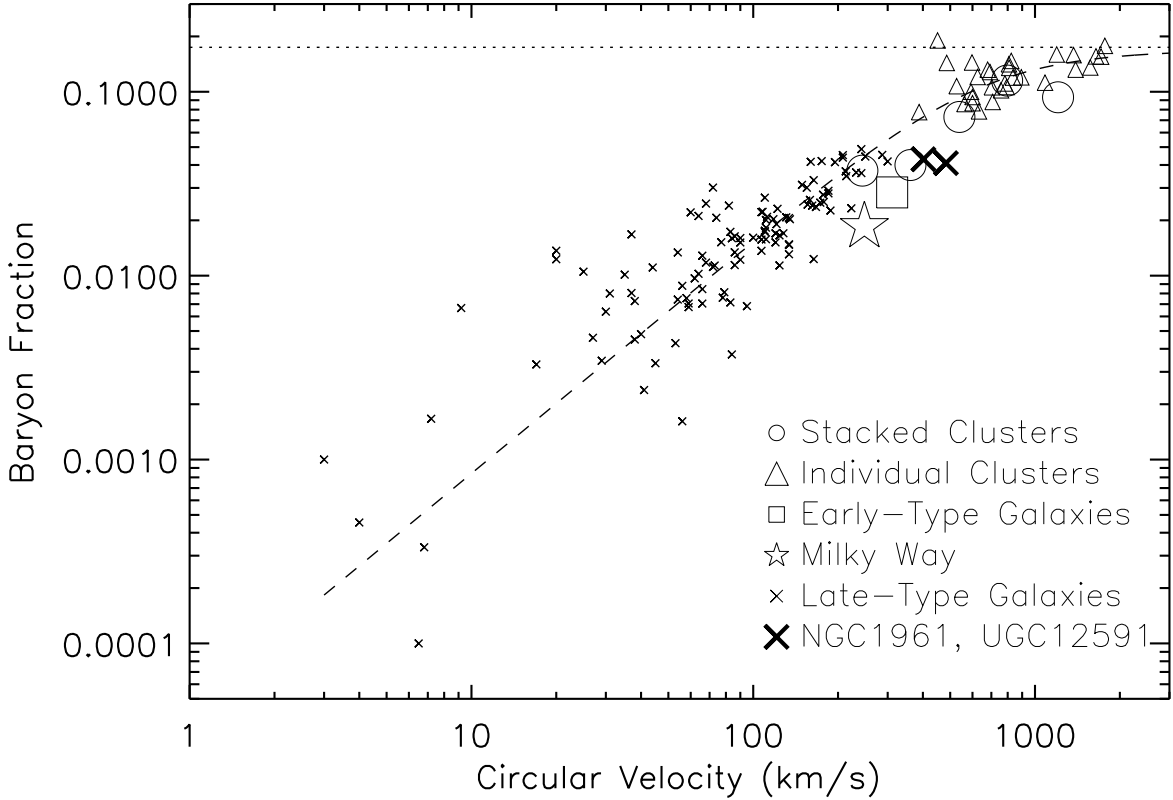}
   \caption{The baryon fraction as a function of the gravitational potential well indicated by the circular velocity at \rtwo.  We plot the new measurements from massive spiral galaxies \ugc\ and NGC~1961 (Anderson \& Bregman 2011) over the archival data from Sakamoto et al.\ (2003), McGaugh (2005), Flynn et al.\ (2006), Vikhlinin et al.\ (2006), Gavazzi et al.\ (2007), Walker et al.\ (2007), Stark et al.\ (2009), Sun et al.\ (2009), Dai et al.\ (2010). The dotted line is the cosmological baryon fraction measured from CMB, and the dashed line is our best fit broken power law model for baryon losses.  For massive spiral galaxies like \ugc\ and NGC~1961, the baryon loss is at least 75\%. \label{fig:bf}}
\end{figure}
We plot the baryon fractions of \ugc\ and NGC~1961 against their rotational velocities, a proxy for the depth of the gravitational potential well, in Figure~\ref{fig:bf}.
We include the data composed by Dai et al.\ (2010) including the stacked results from optically-selected clusters (Dai et al.\ 2007; 2010), individual galaxy clusters (Vikhlinin et al.\ 2006; Sun et al.\ 2008) corrected for the stellar component (Dai et al.\ 2010), individual galaxies (McGaugh 2005; Walker et al.\ 2007; Stark et al.\ 2009), elliptical galaxies (Gavazzi et al.\ 2007) and the Milky Way (Sakamoto et al.\ 2003; Flynn et al.\ 2006), and the recent addition of gas-rich late-type galaxies (Begun et al.\ 2008; Trachternach et al.\ 2009) composed by McGaugh (2011).
This enables us to compare the baryon loss across a large range of systems from dwarf galaxies to rich galaxy clusters.  
For massive spiral galaxies such as \ugc\ and NGC~1961, their dark matter halo masses are close to that of a medium galaxy group, and we find that their baryon losses are comparable to the stacked results for galaxy groups in Figure~\ref{fig:bf}.
For different systems as spiral galaxies and galaxy groups, the consistency between their baryon fractions suggests that the baryon loss is ultimately controlled by the potential well of the dark matter halo.
However, currently we still lack constraints from individual galaxy groups to confirm the stacking results of Dai et al.\ (2010).

The overall baryon fractions for all systems can be fit by a broken power law model (Dai et al.\ 2010).
With the addition of new measurements, especially from those gas-rich late-type galaxies (Begun et al.\ 2008; Trachternach et al.\ 2009), we can better fit the power law slope for baryon fractions in less massive systems.
Thus, we re-fit the data with a broken power law model to find that
\begin{equation}
   f_{b} = \frac{0.16 (v_c/643~{\rm km/s})^{a}}{{(1+(v_c/643~{\rm km/s})^{c})}^{b/c}},
\end{equation}
where $a=1.26$, $b=1.24$, and $c=2$.  The baryon fraction, $f_{b}$, scales as $f_b\propto v_c^{a-b=0.02}$ above the break and  $f_{b} \propto v_c^{a=1.26}$ below the break, 
and the parameter $c$ in the equation is the smoothness of the broken power law model, which is fixed in our fit.
Comparing with the fit in Dai et al.\ (2010), we find the major difference lies in the shallower slope $f_b \propto v_c^{1.26}$ for the baryon loss in galaxies.
We also find a larger break location and a shallower slope for galaxy clusters.

\subsection{Cooling of the Gas Halo}
The gas halo is predicted to play an important role in galaxy formation and evolution.
With our detection of the gas halo emission in \ugc, we can estimate the cooling time of this hot halo and the implied accretion rate onto the galaxy, which can provide constraints on the gas available for new star formation. 
We define the cooling radius as the radius where the cooling time is 10~Gyr, using the expression of the cooling time (Fukugita \& Peebles 2006),
\begin{equation}
\tau(r) = \frac{1.5 nkT}{\Lambda n_e \left(n-n_e\right)} \approx \frac{1.5kT\times1.92}{\Lambda n_e \times 0.92},
\end{equation}
where the latter expression assumes a primeval He abundance resulting in a total particle density of $n = 1.92 n_e$. 
For $T = 10^{6.85}$~K, $Z/ Z_{\odot} = 0.5$, and $\Lambda = 10^{-22.85}$ erg cm$^{3}$ s$^{-1}$ (Sutherland \& Dopita 1993), the cooling radius is at $n_e = 6.8\times10^{-4}$ cm$^{-3}$. For the range of best-fit $\beta$-model profiles constrained in this paper, this corresponds to a cooling radius between 15.6 and 18.0 kpc, and a hot halo mass of $6.2-9.2\times10^8 M_{\odot}$ within that radius. We can roughly estimate the cooling time and rate by dividing the thermal energy in the hot gas within the cooling radius by the luminosity within that radius, and this yields a wide range in cooling time of 2.8-6.3 Gyr for material within the cooling radius, but a fairly narrow range in the effective cooling rate of $0.15-0.21 M_{\odot}$ year$^{-1}$. 
This halo accretion rate is two orders of magnitude too low to assemble the stellar mass of this galaxy within a Hubble time. 
Therefore, significant accretion must have occurred via some other mode, such as cold flows or mergers, to produce the stellar mass seen in this galaxy today, confirming the conclusion drew in NGC~1961 (Anderson \& Bregman 2011).

\acknowledgements

\end{document}